\documentclass[new]{JHEP3}
\usepackage{axodraw} 
\usepackage{amsmath}

\newcommand{\ii}{\'\i} 
 
\def\beq{\begin{eqnarray}}    %%%  begequation/eqnarray
\def\eeq{\end{eqnarray}}      %%%  endequation/eqnarray
\def\tr{\,\mbox{tr}\,}                  %%% trace
\def\Tr{\,\mbox{Tr}\,}                  %%% Trace
\def\al{\alpha}
\def\be{\beta}

\def\de{\delta}

\def\La{\Lambda}
\def\la{\lambda}
\def\na{\nabla}
\def\pa{\partial}

\def\om{\omega}
\def\ph{\varphi}

\def\th{\theta}

\def\Ga{\Gamma}
\def\De{\Delta}
\def\La{\Lambda}

\def\Om{\Omega}

\title{Renormalization Group and Decoupling in Curved Space}

\author{E. V. Gorbar 
\\
Departamento de F\ii sica, 
ICE, Universidade Federal de Juiz de Fora
\\ 
MG, Brazil and 
\\
Bogolyubov Institute for Theoretical Physics, Kiev, Ukraine
\\
E-mail: \email{gorbar@fisica.ufjf.br}}

\author{Ilya L. Shapiro\thanks{On leave from Tomsk State 
Pedagogical University, Tomsk, Russia.}
\\
Departamento de F\ii sica, ICE, Universidade Federal de 
Juiz de Fora
\\ 	
MG, Brazil
\\ 
E-mail: \email{shapiro@dftuz.unizar.es}}

\abstract{It is well known that the renormalization group equations
depend on the scale where they are applied. This phenomenon
is especially relevant for the massive fields in curved 
space, because the decoupling effects may be responsible
for important cosmological applications like the graceful 
exit from the inflation and low-energy quantum dynamics of 
the cosmological constant. We investigate, using both 
covariant and non-covariant methods of calculations and 
mass-dependent renormalization scheme, the vacuum quantum 
effects of a massive scalar field in curved space-time. 
In the higher derivative sector we arrive at the explicit 
form of decoupling and obtain the $\beta$-functions in both 
UV and IR regimes as the limits of general expressions. 
For the cosmological and Newton constants the corresponding 
$\beta$-functions are not accessible in the perturbative 
regime and in particular the form of decoupling remains 
unclear.}

\keywords{Renormalization Group, Physics of the Early Universe}
\received{October 29, 2002}
\accepted{?}
\preprint{DF/UFJF-02/02}

\begin{document}
%%%%%%%%%%%%%%%%%%%%%%%%%%%%%%%%%%%%%%%%%%%%%%%%%%%%%%%%%%%%%%%

$\,$

$\,$

$\,$

\section{Introduction}

The renormalization group methods represent one of the most
universal tools in the Quantum Field Theory and beyond: 
from string theory to lattice QCD and statistical mechanics. 
In the last decades, there was a growing interest to the 
renormalization group for quantum fields in curved 
space-time, especially in relation to the anomalies and 
cosmological applications. The renormalization group and 
related phenomena of the decoupling of the massive quantum
matter fields in an external gravitational field are especially 
relevant for such phenomena as the graceful exit from 
the anomaly-induced inflation \cite{insusy,shocom} and 
the possible low-energy quantum dynamics of the cosmological 
constant \cite{nova}. The relevance of these applications 
(which may help in better understanding of the coincidence 
puzzle for the {\sl $\La$CDM} cosmological model and also 
may lead to a natural theory of inflation) requires 
a detailed quantitative analysis of the physical aspects of 
the renormalization group. In this paper we are going to 
start this investigation.

The standard approach to the renormalization group in an 
external gravitational field has been formulated in 
\cite{nelspan,tmf} (see also 
\cite{buchodin,parker}) on the basis of the Minimal 
Subtraction (MS)
renormalization scheme or the modified Minimal Subtraction 
scheme $\overline{\rm MS}$. In the framework of the 
$\overline{\rm MS}$ scheme, 
the renormalization of quantum matter in curved space-time is 
well understood \cite{book}. The principal results are quite 
simple and look as follows. 
Starting from the renormalizable QFT in the 
flat space-time, one can always construct the renormalizable 
QFT in curved space-time. In particular, if we treat the 
gravitational field as a perturbation over the flat background, 
then the presence 
of an external gravity does not increase the superficial degree 
of divergence for a given Feynman diagram. Taking the 
covariance and locality considerations into account, one can 
easily determine the complete set of possible divergent 
structures and identify the corresponding necessary counterterms. 
In general, the action of the renormalizable theory includes 
the covariant generalization
of the action in flat space plus a set of curvature-dependent 
local terms, which are necessary for the renormalizability. 
In particular, one has to introduce a 
special action of vacuum (external gravity\footnote{Let us 
emphasize that gravity itself is not quantized in this 
paper, so we are confined to the so-called semiclassical 
approach.}) which has the form
\beq
S_{vacuum}\, =\,S_{EH} \, + \,S_{HD}\,.
\label{vacuum}
\eeq
Here the first term is the Einstein-Hilbert action 
with cosmological constant.
\footnote{
Our notations are Euclidean metric 
$\eta_{\mu\nu}=diag(++++)$ and the definition of the 
curvature tensor
$\,\,R_{\mu\nu}=\partial_\lambda\,\Ga^\la_{\mu\nu}
 -\,...\,\,.\,\,$ }
\beq
S_{EH}\, =\,
-\,\frac{1}{16\pi G}\,\int d^4x\,g^{1/2}\,(R + 2\La)\,.
\label{Einstein}
\eeq
The second action contains necessary higher derivative terms
\beq
S_{HD}\, =\, \int d^4x\,g^{1/2}\,\left\{\, 
a_1 C^2 + a_2 R^2 + a_3 E + a_4 {\nabla^2} R \,\right\}\,,
\label{higher}
\eeq
where 
\beq
C^2=R_{\mu\nu\al\be}^2 - 2R_{\al\be}^2 + 1/3\,R^2
\label{vacuum 0}
\eeq
is the square of the Weyl tensor and 
\beq
E = R_{\mu\nu\al\be}R^{\mu\nu\al\be}
-4 \,R_{\al\be}R^{\al\be} + R^2
\label{vacuum 4}
\eeq
is the integrand of the Gauss-Bonnet topological invariant.
$\,a_1,...,a_4$ (and also $\,G\,$ and $\,\La\,$ in (\ref{Einstein}))
are the parameters of the vacuum action. In the present paper 
we shall focus our attention on the renormalization of the 
vacuum parameters $\,a_i$, $\,G$, $\,\La\,$ and on
the quantum corrections to the vacuum action (\ref{vacuum}). 

For the interacting fields there is also renormalization in the 
matter fields sector. Let us remark that the renormalization 
of those terms that have direct analogs in flat space is exactly 
the same as in the flat space \cite{book}. In addition, there is
also the renormalization of the nonminimal parameter
$\xi$ in the scalar action 
\beq
S_s\,=\,\frac12\,\int d^4x\,g^{1/2}\,\left\{
g^{\mu\nu}\pa_\mu\ph\pa_\nu\ph + m^2\ph^2 + \xi R\ph^2 \right\}\,.
\label{scalar}
\eeq
The renormalization of $\xi$ has many interesting features
(see, e.g. \cite{book,hath,brv}) but is beyond the scope of the 
present paper because, in the cosmological framework, 
the effect of this renormalization is less important than 
the renormalization of vacuum terms.

The vacuum divergences of the theory (\ref{scalar}) can 
be easily calculated using the Schwinger-DeWitt method and 
dimensional regularization
$$
{\bar \Ga}^{(1)}_{div}
\,=\,-\,\frac{(\mu^2)^{\om-2}}{2(4\pi)^2(\om-2)}
\,\int d^{2\om}x\,g^{1/2}\,\left\{
\frac{1}{180}(R_{\mu\nu\al\be}^2-R_{\al\be}^2-\na^2R)
-\frac16\Big(\xi-\frac16\Big)\na^2R+
\right.
$$
\beq
\left.
+\frac12\,\Big(\xi-\frac16\Big)^2R^2
+ m^2\Big(\xi-\frac16\Big)\,R +\frac12\,m^4\right\}\,,
\label{scalar 23}
\eeq
where $\,\om\,$ is the parameter of the dimensional 
regularization ($2\om\,$ is the space-time dimension)
and $\mu$ is a dimensional parameter which is introduced 
in order to compensate the change of the dimension.
The renormalization of the vacuum action in the 
$\overline{\rm MS}$ scheme 
proceeds as follows. One has to introduce the local 
counterterm $\,\De S \,$
and then compare the bare $S_0[\La^{(0)},G^{(0)},a^{(0)}_i]$ 
and renormalized  $S_R[\La,G,a_i]=S+\De S$ actions. This 
leads to the standard renormalization relations (see, e.g. 
\cite{book}). In particular,
\beq
-\,\frac{\La^{(0)}}{8\pi G^{(0)}}\,=\,(\mu^2)^{\om-2}\,
\left[-\,\frac{\La}{8\pi G}
\,+\,\frac{m^4}{4\,(4\pi)^2 (\om-2)}\right]\,.
\label{renormalize}
\eeq
Taking into account the fact that $\,\La^{(0)}\,$ does not 
depend on $\mu$, we arrive at the usual (see, e.g. \cite{Brown}) 
renormalization group equation for the cosmological constant
\beq
\mu\,\frac{d}{d\mu}\,\left( \frac{\La}{8\pi G} \right)
\,= \,\be_\La(\overline{MS})\,= \,\frac{m^4}{2(4\pi)^2}\,,
\label{group}
\eeq
where we have set $\om=2$. Using similar
calculations, we arrive at the renormalization group equations
for other vacuum parameters (see, e.g. \cite{nelspan})
$$
\mu\,\frac{d}{d\mu}\,\left(-\frac{1}{16\pi G}\right)
\,= \,\be_R(\overline{MS})
\,= \,\frac{m^2}{(4\pi)^2}\,\Big(\xi-\frac16\Big)\,,
$$
$$
\mu\,\frac{da_1}{d\mu}\,= \,\be_1(\overline{MS})
\,= \,-\,\frac{1}{120\,(4\pi)^2}\,,
$$
\beq
\mu\,\frac{d\,a_2}{d\mu}
\,=\,\be_2(\overline{MS})
\,=\,-\,\frac{1}{2\,(4\pi)^2}\,\Big(\xi-\frac16\Big)^2\,.
\label{group1}
\eeq
Let us emphasize that these equations were obtained within 
covariant formalism and without imposing restrictions on the
external metric $g_{\mu\nu}$.

The interpretation of the renormalization group equations in 
curved space-time requires a special attention. Of course, this 
is a usual situation for the $\overline{\rm MS}$ scheme, where 
even in flat space one has to make a special effort in order 
to interpret the parameter $\mu$.
Since $\mu$ is an auxiliary parameter introduced in 
the dimensional regularization, to get physical results, one 
should trade it for some physical quantity.
This procedure may lead to ambiguities, which can be resolved
in the framework of the more physical mass-dependent scheme of 
renormalization. For the usual flat-space theories, say 
in QED or Yang-Mills, the interpretation of $\mu$ can be 
achieved as follows.
One can compare the renormalization group in the 
$\overline{\rm MS}$ scheme and in a mass-dependent scheme,
which must coincide in the high energy limit. 
In this way, $\mu$ acquires a natural interpretation. 
Depending on the situation
this may be the energy of scattering of quanta,
VEV of the scalar field or temperature etc. 

On the other hand, at low energies, the physically correct
result requires the use of a mass-dependent scheme. But, in the 
case of the vacuum renormalization in a curved space-time, the 
notion of energy for the gravitational quanta is not well 
defined, and the relation between $\overline{\rm MS}$ and 
mass-dependent scheme is a priori unclear. At the same time, 
there is a considerable interest in implementing ideas of 
effective Quantum Field Theory into the gravitational framework: 
both for the classical gravitational background
and effective quantum gravity \cite{don}.
To this end, we need to achieve a quantitative description of
how the decoupling of the heavy particles in the external 
gravitational field occurs.

Let us briefly remind how decoupling proceeds in flat space
in a mass-dependent scheme by using the QED example 
(see, e.g. \cite{manohar}). The 1-loop vacuum polarization 
in the dimensional regularization is equal to
\beq
\frac{e^2}{2\pi^2}\,(p_{\mu}p_{\nu}-p^2g_{\mu\nu})\, 
\Big[\frac{1}{6(\om -2)} -
\frac{\gamma}{6} - \int_0^1 dx \, x(1-x) \,\, 
\ln \, \frac{m_e^2+p^2x(1-x)}{4\pi \mu^2}
\Big]\,,
\label{polari}
\eeq
where $p$ is an external momentum, $m$ is fermion mass, 
$\gamma$ is Euler`s constant, and $\mu$ is the dimensional 
parameter of the regularization. By subtracting in the
$\overline{MS}$ scheme the $\,\frac{1}{\om-2}\,$-pole and the
$\gamma$-term, one find the following contribution:
\beq
-\,\frac{e^2}{2\pi^2}\,(p_{\mu}p_{\nu}-p^2g_{\mu\nu})\, 
\,\int_0^1 dx \, x(1-x) \,\, 
\ln\,\frac{m_e^2+p^2x(1-x)}{4\pi \mu^2}\,.
\label{polari-fin}
\eeq
The $\beta$-function in the $\overline{\rm MS}$ scheme is obtained by 
calculating the derivative $(e/2)\mu d/d\mu$ 
acting on the coefficient of the operator 
$\,\,(p_{\mu}p_{\nu}-p^2g_{\mu\nu})$. The result is 
well known
\beq
\beta_e(\overline{\rm MS})\,\,=\,\,\frac{e^3}{12\pi^2}.
\label{betaMS}
\eeq
Thus, the $\beta$-function in the $\overline{MS}$ scheme is a 
constant independent of the fermion mass and $\mu$.
To calculate the $\beta$-function in a mass-dependent scheme 
such as, e.g., the momentum space subtraction scheme, one has 
to subtract the value of the graph at a point $p^2=M^2$ and 
calculate the derivative $(e/2)M d/dM$ acting on the coefficient 
of the $\,\,(p_{\mu}p_{\nu}-p^2g_{\mu\nu})$ projector. 
As a result, we obtain
\beq
\beta_e= \frac{e^3}{2\pi^2}\int_0^1 \, dx\, x(1-x)\, 
\frac{M^2x(1-x)}{m_e^2+M^2x(1-x)}.
\label{beta-md}
\eeq
It is easy to verify that when the renormalization point is much 
larger than the fermion mass, the $\beta$-function obtained in 
the mass-dependent scheme coincides with the $\overline{MS}$ 
$\,\,\beta$-function (\ref{betaMS}). As the renormalization 
point passes through $m_e$, massive fermion decouples and, 
for $\,\,M \ll m_e$, its contribution to the $\beta$-function 
vanishes as
\beq
\beta_e(M \ll \mu_e)\, 
\approx\, \frac{e^3}{60\pi^2}\frac{M^2}{m_e^2}.
\label{lowenergy}
\eeq
Thus, the $\overline{MS}$ $\beta$-function gives wrong result at 
energies less than $m_e$. 
Decoupling of heavy particles is implemented in 
the $\overline{MS}$ scheme by putting, artificially, the 
$\beta$-function to zero for $\mu < m$. This can be called 
``hard decoupling'' or ``sharp cut-off'' approximation. 
Indeed, this is 
just an approximation to the real picture of decoupling,
which may be essentially scheme-dependent. In particular, 
the example of QED shows that, in principle, the decoupling 
in gravity can proceed in a soft manner such that the 
$\beta$-function is continuous at the scale $\,\,m$.

The purpose of this paper is to perform, for the first time,
the derivation of the renormalization group equations for the
parameters $\,\La$, $\,G$, $\,a_1\,$ and  $\,a_2\,$ in a 
simple mass-dependent scheme. We consider a massive 
non-minimally coupled scalar theory, but the results can be 
also generalized for massive fermions and vectors. 
In section 2 we present the derivation of 
the polarization operator of gravitons using dimensional 
regularization. The table of cumbersome integrals are
exposed in the Appendix A. Although we will later 
confirm our principal conclusions by using covariant heat 
kernel calculations, we believe that the linearized gravity 
calculations are useful. These calculations are performed on 
the flat space background and, hence, they are 
maximally close to usual perturbative quantum field theory 
calculations, where we have a good intuition of how 
the decoupling of massive particles works.
In section 3, the polarization operator of section 2 is 
verified using the previously known expressions for the 
summation of the Schwinger-DeWitt series \cite{Avramidi,bavi2}. 
In section 4, 
the renormalization group equations for the vacuum parameters 
are derived and the comparison with similar equations 
in the $\overline{\rm MS}$ scheme is given. Another 
calculation which is aimed to clarify the relation between
the $\overline{\rm MS}$ and mass-dependent scheme is presented  
in the Appendix 
B, where we apply the covariant cut-off regularization of the 
proper-time integral representation for the effective action.

Unfortunately, in this regularization scheme one can recover 
the reliable form of the $\be$-functions only at high energy
limit, while the small cut-off limit is inconsistent.
In section 5 we discuss the results of the calculations in 
the mass-dependent scheme. In particular, this opportunity 
is used to discuss the form of the effective action
which could correspond to the running of the cosmological 
and inverse Newton constants at low energies. 
After all, we draw our conclusions. 

In order to avoid the confusion, let us fix the notations 
for the distinct types of the $\,\be$-functions. We shall 
denote, in this paper, the $\overline{\rm MS}$-scheme 
$\,\be$-function for the effective charge $\,C\,$ as
$\,\be_C(\overline{\rm MS})$ and the $\,\be$-function for 
the same object, derived in a mass-dependent scheme as
$\,\be_C$. The high energy limit of this $\,\be$-function
will be denoted as $\,\be^{UV}_C\,$ and the low-energy
limit as $\,\be^{IR}_C$. Of course, we expect that the 
correctly defined $\,\be$-function would satisfy the 
relation
$$
\be^{UV}_C\,\,=\, \,\be_C(\overline{\rm MS})
\,+\, {\cal O} \Big(\frac{m^2}{p^2}\Big)\,.
$$

%%%%%%%%%%%%%%%%%%%%%%% / * \ %%%%%%%%%%%%%%%%%%%%%%%%%
 %%%%%%%%%%%%%%%%%%%%% /*%#%*\ %%%%%%%%%%%%%%%%%%%%%%%
 %%%%%%%%%%%%%%%%%%%%% \*%#%*/ %%%%%%%%%%%%%%%%%%%%%%%
%%%%%%%%%%%%%%%%%%%%%%% \ * / %%%%%%%%%%%%%%%%%%%%%%%%%
%%% 2
\section{The non-covariant perturbative calculation}
\label{sec.2} 

The renormalization group for the parameters of the 
vacuum action can be established through the perturbative 
calculation of the quantum corrections to the gravitational 
2-point function. Therefore, in this section we shall 
derive the contributions of the loop of massive particle 
to the propagator of the gravitational
perturbations $h_{\mu\nu}$ on the flat background 
\beq
g_{\mu\nu}=\eta_{\mu\nu}+h_{\mu\nu}\,.
\label{expand}
\eeq 
The calculations will be performed in the Euclidean space.
The diagrams which contribute to the gravitational 
polarization operators are presented at Figure 1. 

%%%%%%%%%%%%%%%%%%%%%%%%%%%%%%%%%%%%%%%%%%%%%%%%%%%%%%%%

 \vskip 1mm
 %% \noindent
%%%%%%%%%%%%%%%%%%%%%%%%%%%%%
 \begin{picture}(120,100)(0,0)
 \Photon(0,25)(25,25){2}{6}
 \BCirc(50,25){25}
 \Photon(75,25)(100,25){2}{6}
 \Vertex(75,25){3} \Vertex(25,25){3}  
 %% \Text(5,15)[c]{$p,\,\al\be$}
 %% \Text(10,15)[c]{$p,\,\nu\mu$}
 \Text(100,60)[c]{(a)}
 \end{picture}
 \begin{picture}(20,100)(0,0) \end{picture}
%%%%%%%%%%%%%%%%%%%%%%%%%%%%%
 \begin{picture}(120,100)(0,0)
 \Photon(0,0)(100,0){2}{24}  
 \BCirc(50,25){25}
 %% \Text(10,15)[c]{$p,\,\al\be$}
 %% \Text(10,15)[c]{$p,\,\nu\mu$}
 \Vertex(50,0){3}  
 %% \Text(110,25)[c]{(b)}
 \Text(100,60)[c]{(b)}
 \end{picture}
 \begin{picture}(20,100)(0,0) \end{picture}
%%%%%%%%%%%%%%%%%%%%%%%%%%%%%
\begin{picture}(120,100)(0,0)
 \Photon(0,0)(100,0){2}{24}  
 \Photon(50,0)(50,25){2}{6}  
 \BCirc(50,50){25}
 %% \Text(10,15)[c]{$p,\,\al\be$}
 %% \Text(10,15)[c]{$p,\,\nu\mu$}
 \Vertex(50,0){3}  
 \Vertex(50,25){3}  
 \Text(100,60)[c]{(c)}
 %% \Text(110,25)[c]{(c)}
 \end{picture}
%%%%%%%%%%%%%%%%%%%%%%%%%%%%%%
 \vskip 3mm
 \begin{quotation}
 \noindent
 {\small\it Figure 1. $\,\,$  
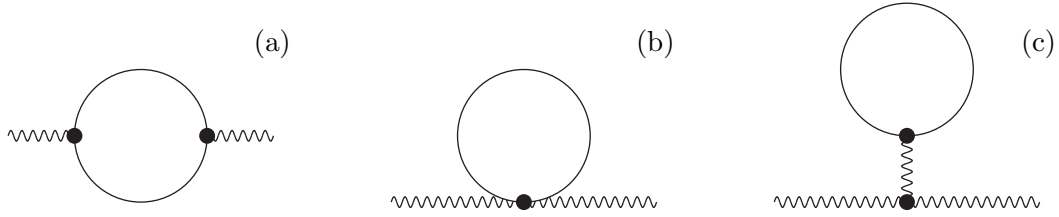
 The $1$-loop diagrams with the loop of massive 
 scalar and two external gravitational lines.
 The straight lines correspond to the massive scalar and wavy 
 lines to the external field $h_{\mu\nu}$. }
 \end{quotation}
 \vskip 1mm

The $3$-point and $4$-point vertices (see Figure 2) have the 
form 
\beq
V^{\al\be}(k,p,q)=(2\pi)^4\Big[
\frac12(\eta^{\al\be}p\cdot q - q^\al p^\be - p^\al q^\be)
+\xi(k^\al k^\be - k^2\eta^{\al\be}) 
- \frac{m^2}{2}\eta^{\al\be} \Big];
\label{3}
\eeq
\vskip 2mm
%%%%%%%%%%%%%%%%%%%%%%%%%%%%%%%%%%%%%%%%%%%%%%%%%%%%%%%%%%%%%%%%
$$
V^{\mu\nu\,,\,\al\be}(k,p,q,l)\,=\,(2\pi)^4\,\left\{\,
\de^{\mu\nu\,,\,\al\be}\,\Big[\,\xi\,(k^2+l^2+3 k\cdot l)
-\frac12\,p\cdot q\Big] +
\right.
$$ $$
\left.
+ \eta^{\mu\nu}\eta^{\al\be}\,
\Big[\,\frac14\,p\cdot q - \frac{\xi}{2}\,(k^2+l^2+k\cdot l)
\,\Big]
\,+\, \eta^{\mu\nu}\,\Big[\,-\frac12\,p^\al q^\be 
+ \xi(k^\al k^\be + \frac12\,l^\al l^\be + k^\al l^\be )
\,\Big]+
\right.
$$ $$
\left.
+ \eta^{\al\be}\,\Big[\,-\frac12\,p^\mu q^\nu 
+ \xi(l^\mu l^\nu + \frac12\,k^\mu k^\nu + k^\mu l^\nu )
\,\Big]
\right.
$$
\beq
\left.
+ \eta^{\mu\al}\,\Big[\,p^\nu q^\be + q^\nu p^\be
- 2\xi\,(k^\nu k^\be + l^\nu l^\be )
- 2\xi\,k^\nu l^\be - \xi l^\nu k^\be 
\,\Big]\,\right\}\,,
\label{4}
\eeq
where, for the sake of compactness, we do not maintain the 
symmetry inside the couples of indices 
$(\al\be)$ and $(\mu\nu)$.

%%%%%%%%%%%%%%%%%%%%%%%%%%%%%%%%%%%%%%%%%%%%%%%%%%%%%%%%
%%%% \newpage
%% \vskip 5mm
%% \centerline{\Large\bf Place for Figure 2}
%% \vskip 5mm
 \vskip 1mm
%%%%%%%%%%%%%%%%%%%%%%%%%%%%%
 \begin{picture}(20,80)(0,0) \end{picture}
 \begin{picture}(120,80)(0,0)
 \Photon(20,25)(45,25){2}{8}
 \Line(45,25)(60,0)  
 \Vertex(45,25){3}
 \Line(45,25)(60,50)
 \Text(10,15)[c]{$k,\,\al\be$}
 \Text(70,5)[c]{$p$}
 \Text(70,55)[c]{$q$}
 \end{picture}
 \begin{picture}(20,80)(0,0) \end{picture}
 \begin{picture}(20,80)(0,0) \end{picture}
%%%%%%%%%%%%%%%%%%%%%%%%%%%%%
 \begin{picture}(120,80)(0,0)
 \Photon(20,0)(45,25){2}{8}
 \Photon(20,50)(45,25){2}{8}
 \Vertex(45,25){3}
 \Line(45,25)(70,0)  
 \Line(45,25)(70,50)
 \Text(5,5)[c]{$k,\,\mu\nu$}
 \Text(5,55)[c]{$l,\,\al\be$}
 \Text(85,5)[c]{$p$}
 \Text(85,55)[c]{$q$}
 \end{picture}
 %%%%%%%%%%%%%%%%%%%%%%%%%%%%%%

 \vskip 3mm
 \begin{quotation}
 \noindent
 {\small\it Figure 2. $\,\,$  
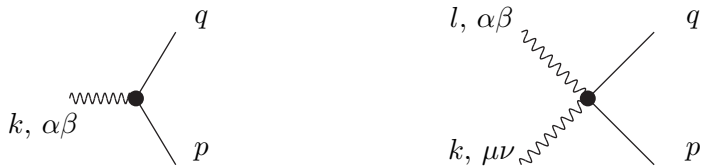
 Two relevant vertices of the gravity-scalar interactions.}
 \end{quotation}
 \vskip 1mm

The first observation is that the diagram ($1c$) does not 
contribute to the renormalization group because we are 
dealing with the semiclassical approximation. The object 
of interest is the effective action of the external metric, 
and its expansion into series in $h_{\mu\nu}$ does not lead 
to the diagram ($1c$). For this reason we have to consider 
only ($1a$) and ($1b$) diagrams. Let us start from ($1a$).
The lengthy calculations in the dimensional regularization
give the following expression for the polarization operator
$$
{\Pi}^{\mu\nu\,,\,\al\be}_{1a}(p)
\,=\,-\, \frac{1}{16}\,\eta^{\mu\nu}\eta^{\al\be}\,
(p^4I_1 -2p^4I_2+6p^2I_3+2p^4I_4+8I_9)
\,-\,\frac12\,p^\mu p^\nu p^\al p^\be \,(I_7-2I_5+I_4)\,-
$$
$$
\,-\,\frac18\,(\eta^{\mu\al}p^\nu p^\be+\eta^{\mu\be}p^\nu p^\al
+\eta^{\nu\al}p^\mu p^\be+\eta^{\nu\be}p^\mu p^\al)
\,(4I_8-4I_6+I_3)\,-\,\de^{\mu\nu\,,\,\al\be}\,I_9\,-
$$
$$
\,-\,\frac18\,(\,\eta^{\mu\nu}p^\al p^\be + \eta^{\al\be}p^\mu p^\nu\,)
\,(p^2I_4 -p^2I_2+4I_8-4I_6\,)
\,-\,\frac{\xi^2}{2} \,I_1\,(p^\mu p^\nu -p^2\eta^{\mu\nu})\,
(p^\al p^\be -p^2\eta^{\al\be})\,-
$$
$$
-\,\frac{\xi}{8}\,(p^\mu p^\nu -p^2\eta^{\mu\nu})\,
\eta^{\al\be}(p^2I_1 + 4I_3)
\,-\,\frac{\xi}{8}\,(p^\al p^\be -p^2\eta^{\al\be})\,
\eta^{\mu\nu}(p^2I_1 + 4I_3)\,+
$$
\beq
+\,\frac{\xi}{4}\,(p^\mu p^\nu -p^2\eta^{\mu\nu})\,
p^\al p^\be (I_1 -2 I_4)
\,+\,\frac{\xi}{4}\,(p^\al p^\be -p^2\eta^{\al\be})\,
p^\al p^\be (I_1 -2 I_4)\,+
\label{fish}
\eeq
$$
+\,\frac{{\Ga}(1-\om)}{2(4\pi)^\om}\,(m^2)^{\om-1}\,
\Big[\,\frac{\xi}{2}\, (p^\mu p^\nu \eta^{\al\be} 
+ p^\al p^\be \eta^{\mu\nu})
\,+\, \Big(\frac14 - \xi\Big)\, 
p^2 \eta^{\al\be} \eta^{\mu\nu}\,\Big]
+\,\frac{{\Ga}(-\om)}{4(4\pi)^\om}\,(m^2)^\om\,
\eta^{\mu\nu}\eta^{\al\be}\,,
$$
where the integrals $I_1$, ..., $I_9$ are defined in the
Appendix A.
\vskip 2mm

The second diagram is simpler and the corresponding
expression is
$$
{\Pi}^{\mu\nu\,,\,\al\be}_{1b}(p)\,=\,
\frac{\xi}{4}\,\frac{{\Ga}(1-\om)}{(4\pi)^\om}\,(m^2)^{\om-1}\,
\Big[\,p^2\de^{\mu\nu\,,\,\al\be}
\,-\,p^2\eta^{\mu\nu}\,\eta^{\al\be}\,+
$$
$$
+\,(p^\mu p^\nu \eta^{\al\be}+p^\al p^\be \eta^{\mu\nu})
\,-\,
\frac12\,(\eta^{\mu\al}p^\nu p^\be+\eta^{\mu\be}p^\nu p^\al
+\eta^{\nu\al}p^\mu p^\be+\eta^{\nu\be}p^\mu p^\al)\,\Big]\,+
$$
\beq
+\,\frac{{\Ga}(-\om)\,(\om-2)}{8(4\pi)^w}\,(m^2)^\om\,
\Big(\,2\de^{\mu\nu\,,\,\al\be}-\eta^{\mu\nu}\,\eta^{\al\be}   
\,\Big)\,.
\label{tadpole1}
\eeq
\vskip 3mm

In order to apply the mass-dependent renormalization scheme,
we need both divergent and finite parts of the polarization
operator. Let us start from divergencies and establish the
correspondence with the covariant expression (\ref{scalar 23}).

For this end, since we are working with the polarization 
operator, the terms in the vacuum action (\ref{vacuum}) must 
be expanded up to the second order in $h_{\mu\nu}$.
As far as the Gauss-Bonnet term does not contribute to the 
propagator, we can replace the $C^2_{\mu\nu\al\be}$-term
by the expression $2W$, where $W=R^2_{\mu\nu}-\frac13\,R^2$.
Then the relevant bilinear expressions have the form
\beq
\int d^4x\,g^{1/2}\,=\,\int d^4x\,h^{\mu\nu}\,
\Big(\,\frac18\,\eta_{\mu\nu}\,\eta_{\al\be}
-\,\frac14\,\de_{\mu\nu\,,\,\al\be}\,\Big)\,h^{\al\be}\,+...\,, 
\label{CC}
\eeq
\vskip 2mm
$$
\int d^4x\,g^{1/2}R\,=\,\int d^4x\,\,\,h^{\mu\nu}\,
\Big[\,\frac14\,\de_{\mu\nu\,,\,\al\be}\pa^2
-\frac14\,\eta_{\mu\nu}\,\eta_{\al\be}\pa^2+
$$
\beq
+\frac14\,
(\eta_{\mu\nu}\pa_\al\pa_\be + \eta_{\al\be}\pa_\mu\pa_\nu)
-\,\frac12\,\eta_{\mu\al}\pa_\nu\pa_\be)
\,\Big]\,h^{\al\be}
\,+...\,, 
\label{R}
\eeq
$$
\int d^4xg^{1/2}R^2=\int d^4x\,\,h^{\mu\nu}
\Big[\,\pa_\mu\pa_\nu \pa_\al \pa_\be
+\eta_{\mu\nu}\eta_{\al\be}\pa^2-
$$
\beq
-(\eta_{\mu\nu}\pa_\al\pa_\be 
+ \eta_{\al\be}\pa_\mu\pa_\nu)\pa^2
\Big]h^{\al\be}
+..\,, 
\label{2R}
\eeq
$$
\int d^4x\,g^{1/2}W\,=\,\int d^4x\,\,\,h^{\mu\nu}\,
\Big[\,\frac14\,\de_{\mu\nu\,,\,\al\be}\,\pa^4
- \frac{1}{12}\,\eta_{\mu\nu}\,\eta_{\al\be}\,\pa^4
+ \frac16\,\pa_\mu\pa_\nu\pa_\al\pa_\be
- \frac12\,\eta_{\mu\al}\,\pa_\be\pa_\nu\,+
$$
\beq
+ \,\frac{1}{12}\,(\eta_{\mu\nu}\pa_\al\pa_\be
+ \eta_{\al\be}\pa_\mu\pa_\nu)\,\pa^2
\,\Big]\,h^{\al\be}\,+...\,, 
\label{W}
\eeq
where the dots stand for the lower and higher order terms. 

Taking into account the formulas
$$
{\Ga}(2-\om)\, = \,\frac{1}{2-\om}\, + \,{\cal O}(2-\om)\,,
$$$$
{\Ga}(1-\om)\,=\,-\,\frac{1}{2-\om}\,-\,1\,+\,{\cal O}(2-\om)\,,
$$
\beq
{\Ga}(-\om)\,=\,\frac{1}{2(2-\om)}\,+\,\frac34\,
+\,{\cal O}(2-\om)\,,
\label{gamma}
\eeq
one can easily identify the divergent part of the 
polarization operator (sum of the two expressions
(\ref{fish}) and (\ref{tadpole1})), with the expansion
(\ref{CC}) - (\ref{W}) of the divergent part of the 
effective action

$$
{\bar \Ga}^{(1)}_{div}(diagr)=-\frac{1}{2(4\pi)^2(\om-2)}
\int d^4x\,g^{1/2}\left\{
\frac{1}{60}W + \frac12\Big(\xi-\frac16\Big)^2R^2\,+
\right.
$$
\beq
\left.
+ \Big(\xi-\frac16\Big)m^2R + \frac12m^4\right\},
\label{diagram}
\eeq
that is almost the same as Eq. (\ref{scalar 23}). 
The main difference is that the expression 
(\ref{diagram}) does not contain surface terms 
$\,\,\int d^4x\,g^{1/2}E\,\,$ and 
$\,\,\int d^4x\,g^{1/2}\na^2 R\,\,$
which do not contribute to the propagator. 

After subtracting divergences (together with the 
$\,\ln[4\pi\mu^2/m^2]\,$ terms) and neglecting the 
${\cal O}(w-2)$ terms, the polarization operator becomes
$$
{\Pi}^{\mu\nu\,,\,\al\be}(p)\,=\,\frac{1}{2(4\pi)^2}\,\left\{\,
\frac12\,
\de^{\mu\nu\,,\,\al\be}
\,\Big[p^4C_3-2p^2m^2C_2+m^4C_1+\frac{p^4}{40}+\frac{p^2m^2}{4} 
+ \frac{3m^4}{4}\Big]+
\right.
$$$$
\left.
+\frac14
\eta^{\mu\nu}\,\eta^{\al\be}\,
\,\Big[\,p^4(C_1+5C_2+C_3)-p^2m^2(3C_1+2C_2)+m^4C_1
-\frac{9p^4}{40}-\frac{p^2m^2}{4}
-\frac{3m^4}{4}\Big]+
\right.
$$$$
\left.
+\frac12\,(p^\mu p^\nu \eta^{\al\be}+p^\al p^\be \eta^{\mu\nu})
\,\Big[p^2(C_2+2C_3)-2m^2C_2+\frac{p^2}{30}+\frac{m^2}{6}\Big]
+2C_3p^\mu p^\nu p^\al p^\be+
\right.
$$$$
\left.
+\frac14\,(\eta^{\mu\al}p^\nu p^\be+\eta^{\mu\be}p^\nu p^\al
+\eta^{\nu\al}p^\mu p^\be+\eta^{\nu\be}p^\mu p^\al)
\Big[p^2(C_2+4C_3)-m^2(C_1+4C_2)-\frac{p^2}{60}-\frac{m^2}{6}
\Big]\,-
\right.
$$
\vskip 0.1mm
$$
\left.
-2\xi^2 C_1\,(p^\mu p^\nu-p^2\eta^{\mu\nu})
\,(p^\al p^\be-p^2\eta^{\al\be})-
\right.
$$
\vskip 0.1mm
$$
\left.
-\frac{\xi}{2}\,(p^\mu p^\nu-p^2\eta^{\mu\nu})
\,\Big[\Big(C_1p^2-2C_1m^2+2p^2C_2-\frac{p^2}{6}\,
\Big)\eta^{\al\be}+4p^\al p^\be C_2\Big]-
\right.
$$
$$
\left.
-\frac{\xi}{2}\,(p^\al p^\be-p^2\eta^{\al\be})
\,\Big[\Big(C_1p^2-2C_1m^2+2p^2C_2-\frac{p^2}{6}\,
\Big)\eta^{\mu\nu}+4p^\mu p^\nu C_2\Big]\,+
\right.
$$
$$
\left.
+\,\frac{\xi m^2}{2}\,
\Big[\,-p^2\de^{\mu\nu\,,\,\al\be}
+p^2\,\eta^{\mu\nu}\,\eta^{\al\be}
-(p^\mu p^\nu \eta^{\al\be}+p^\al p^\be \eta^{\mu\nu})+
\right.
$$
\beq
\left.
\,+\frac12\,(\eta^{\mu\al}p^\nu p^\be+\eta^{\mu\be}p^\nu p^\al
+\eta^{\nu\al}p^\mu p^\be+\eta^{\nu\be}p^\mu p^\al)\,\Big]
\right\}\,,
\label{large}
\eeq
where
$$
C_1=-\int_{0}^{1/2}dy\,\ln \Big[\frac{p^2}{m^2}
\Big(\frac14-y^2\Big)+1\Big]\,,
$$
$$
C_2=\int_{0}^{1/2}dy\,\Big(\frac14-y^2\Big)\,
\ln \Big[\frac{p^2}{m^2}\Big(\frac14-y^2\Big)+1\Big]\,,
$$
\beq
C_3=-\int_{0}^{1/2}dy\,\Big(\frac14-y^2\Big)^2
\ln \Big[\frac{p^2}{m^2}\Big(\frac14-y^2\Big)+1\Big]
\label{integrals}
\eeq
are functions of the ratio $\,\,p^2/m^2$, which can be easily 
calculated
$$
C_1=A\,,\,\,\,\,\,\,\,\,\,\,\,\,\,\,\,\,\,\,\,\,\,\,
C_2=\frac{A}{36a^2}-\frac{A}{4}+\frac{1}{36}\,,
$$
$$
C_3=\frac{A}{5a^4}-\frac{A}{6a^2}+\frac{A}{16}
+\frac{1}{60a^2}-\frac{41}{3600}\,,
$$
where 
\beq
A\,=\,1-\frac{1}{a}\,\ln\,\Big(\frac{2+a}{2-a}\Big)
\,,\,\,\,\,\,\,\,\,\,\,\,\,\,a^2=\frac{4p^2}{p^2+4m^2}\,.
\label{first A}
\eeq

For the terms without $\xi$ we can compare the 
expression (\ref{large}) and the corresponding 
covariant expression 
\beq
S_{vac}^R \, =\, \int d^4x\,g^{1/2}\,\left\{\, 
- \th_G R - \th_\La + \th_1 C^2 + \th_2 R^2 \,\right\}\,.
\label{action}
\eeq
Using the bilinear expansions (\ref{CC}), (\ref{R}),
(\ref{2R}), (\ref{W}), we arrive at the following relations
for $\th_1$, $\th_2$, $\th_G$, $\th_\La$
in the momentum space (the expression
(\ref{action}) must be understood such that all $\th$'s 
act as operators in the bilinear expansions of the 
corresponding covariant terms):
$$
\frac{\th_1 p^4}{2} + \frac{\th_G p^2}{4}
- \frac{\th_\La}{4}\,=\,-\,\frac{1}{4(4\pi)^2}\,\Big[\,
\frac{C_3 p^4}{2}-p^2m^2C_2+\frac{C_1 m^4}{2}
+\frac{p^4}{80}+\frac{p^2m^2}{8}+\frac{3m^4}{8}\,\Big]\,,
$$
$$
-\,\frac{\th_1 p^4}{3}\,+ \,2\,\th_2 p^4 - \frac{\th_G p^2}{2}
+ \frac{\th_\La}{4}\,=
\,\,\,\,\,\,\,\,\,\,\,\,\,\,\,\,\,\,\,\,\,\,\,\,\,\,\,
\,\,\,\,\,\,\,\,\,\,\,\,\,\,\,\,\,\,\,\,\,\,\,\,\,\,\,\,\,
\,\,\,\,\,\,\,\,\,\,\,\,\,\,\,\,\,\,\,\,\,\,\,\,\,\,\,\,\,
\,\,\,\,\,\,\,\,\,\,\,\,\,\,\,\,\,\,\,\,\,\,\,\,\,\,\,\,\,
\,\,\,\,\,\,\,\,\,\,\,\,\,\,\,\,\,\,\,\,\,\,\,\,\,\,\,
$$$$
\,\,=\,-\,\frac{1}{4(4\pi)^2}\,\Big[\,
\frac{(C_1+5C_2+C_3) p^4}{2}-\frac{p^2m^2(3C_1+2C_2)}{2}
+\frac{m^4C_1}{2}-\frac{9p^4}{80}-\frac{p^2m^2}{8}
-\frac{3m^4}{8}\,\Big]\,,
$$$$
\Big(\frac{\th_1}{3}\,-\,2\th_2\Big)\,p^2 + \frac{\th_G}{2}
\,=\,-\,\frac{1}{4(4\pi)^2}\,\Big[\,(2C_3+C_2) p^2 
- 2m^2\,C_2\,+\,\frac{p^2}{30}\,+\,\frac{m^2}{6}\,\Big]\,,
$$
$$
-\,\frac{\th_1p^2}{2}\,-\,\frac{\th_G}{4}\,=\,-\,
\frac{1}{4(4\pi)^2}\,\Big[\,\Big(2C_3+\frac{C_2}{2}\Big) p^2
-\Big(2C_2+\frac{C_1}{2}\Big) m^2 
- \,\frac{p^2}{120}\,-\,\frac{m^2}{12}\,\Big]\,,
$$
\beq
\frac{2\th_1}{3}\,+\,2\th_2 \,=\,-\,\frac{C_3}{(4\pi)^2}
\,.
\label{relation1}
\eeq
Furthermore, it is easy to check that the $\xi$-dependent terms 
can be cast into the form 
$$
-\frac{\xi}{(4\pi)^2}
\left\{\,(p^\mu p^\nu-p^2\eta^{\mu\nu})
\,(p^\al p^\be-p^2\eta^{\al\be})\,
\Big[2\xi A + \frac{4A}{3a^2}-A+\frac19 \Big]\,+
\right.
$$
$$
\left.
+\,\frac{m^2}{4}\,\Big[\,p^2\de^{\mu\nu\,,\,\al\be}
- p^2\,\eta^{\mu\nu}\,\eta^{\al\be}
+ (p^\mu p^\nu \eta^{\al\be}+p^\al p^\be \eta^{\mu\nu})+
\right.
$$
\beq
\left.
\,-\frac12\,(\eta^{\mu\al}p^\nu p^\be+\eta^{\mu\be}p^\nu p^\al
+\eta^{\nu\al}p^\mu p^\be+\eta^{\nu\be}p^\mu p^\al)\,\Big]
\right\}\,.
\label{large plus}
\eeq
This can be easily identified with Eq. (\ref{2R})
and (\ref{R}), 
therefore the $\xi$-dependent terms are associated with 
the $\int R^2$ and $\int m^2R$-terms in the effective action 
and contribute to the $\th_2$ and $\th_G$ coefficients only.

Despite the number of the equations in (\ref{relation1}) 
exceeds the number of the unknowns, all these equations can  
be indeed satisfied. Taking into account the $\xi$-dependent 
terms in $\th_2$, we obtain
$$
\th_\La \,=\, \frac{3m^4}{8\,(4\pi)^2}\,,
$$
$$
\th_G \,=\,\frac{m^2}{2\,(4\pi)^2}\,\Big(\xi-\frac16\Big)\,,
$$
$$
\th_1\,=\,-\,\frac{1}{4\,(4\pi)^2}\,\Big(\frac{8A}{15a^4}
+\frac{2}{45a^2}\Big)\,,
$$
$$
\th_2 \,= \,\frac{A}{4\,(4\pi)^2}\,
\Big[\,-2\Big(\xi-\frac16\Big)^2+\frac13\,\Big(\xi-\frac16\Big)
\Big(1-\frac{4}{a^2}\Big)-\frac{2}{9a^4}+\frac{1}{9a^2}
-\frac{1}{72}\Big]-
$$
\beq
-\frac{1}{216\,(4\pi)^2\,a^2}\,.
\label{theta}
\eeq

The coefficients (\ref{theta}) contain all information 
about the polarization operator. It will be checked, 
using the covariant Avramidi and Barvinsky - Vilkovisky 
technique in the next section. After that we shall 
consider the renormalization group for the coefficients 
of the vacuum action (\ref{vacuum}).

%%%%%%%%%%%%%%%%%%%%%%% / * \ %%%%%%%%%%%%%%%%%%%%%%%%%
 %%%%%%%%%%%%%%%%%%%%% /*%#%*\ %%%%%%%%%%%%%%%%%%%%%%%
 %%%%%%%%%%%%%%%%%%%%% \*%#%*/ %%%%%%%%%%%%%%%%%%%%%%%
%%%%%%%%%%%%%%%%%%%%%%% \ * / %%%%%%%%%%%%%%%%%%%%%%%%%

\section{The covariant derivation of the effective action}
\label{sec.3} 

The calculation of the polarization operator, presented in 
section 2, is rather 
complicated, so it is useful to verify it using a covariant 
method and compare the two results. To this end, we need to 
obtain the covariant effective action up to the second order 
in curvature. The one-loop effective action can 
be presented as the functional trace of the proper-time 
integral of the heat kernel $K(s)$. This representation of 
the effective action has been obtained by Avramidi using 
summation of the Schwinger-DeWitt series \cite{Avramidi} 
and by Barvinsky and Vilkovisky using generalized 
Schwinger-DeWitt technique \cite{bavi2}. 

In the case of a massive scalar field, the formulas of 
\cite{bavi2} must be slightly modified and the one-loop 
Euclidean effective action (up to the second order in curvature) 
reads (compare to the formulas (1.8) and (2.1) of \cite{bavi2})
\beq
{\bar \Ga}^{(1)}\,=\,-\,\frac12\,{\Tr}{\ln}
\Big(\,-{\na}^2{\hat 1}+m^2-{\hat P}+\frac16\,R{\hat 1}\Big)
\,\,=\,\,\,-\,\frac12\,\int_{0}^{\infty}\,
\frac{ds}{s}\,\tr\,K(s)\,,
\label{effective}
\eeq
where ${\hat P}=-(\xi-1/6)R\,$ and
$$
\tr\,K(s)\,=\,\frac{(\mu^2)^{2-w}}{(4\pi s)^w}\,
\int d^4x\,g^{1/2}\,e^{-sm^2}\,{\rm tr}\,\left\{\,{\hat 1}
\,+\,s{\hat P}
+\,s^2\,\Big[ R_{\mu\nu}f_1(-s\na^2)R^{\mu\nu}
\,+
\right.
$$
\beq
\left.
+ Rf_2(-s\na^2)R + {\hat P}f_3(-s\na^2)R
+ {\hat P}f_4(-s\na^2){\hat P}\,\Big]
\right\}\,+\,{\cal O}({\cal R}^3)\,,
\label{heat}
\eeq
where
$$
f_1(\tau)=\frac{f(\tau)-1+\tau/6}{\tau^2}
\,,\,\,\,\,\,\,\,\,\,\,\,\,\,\,\,\,\,\,
f_2(\tau)=\frac{f(\tau)}{288}+\frac{f(\tau)-1}{24\tau}
-\frac{f(\tau)-1+\tau/6}{8\tau^2}\,,
$$
$$
f_3(\tau)=\frac{f(\tau)}{12}
+\frac{f(\tau)-1}{2\tau}
\,,\,\,\,\,\,\,\,\,\,\,\,\,\,\,\,\,\,\,
f_4(\tau)=\frac{f(\tau)}{2}
$$
and 
$$
f(\tau)=\int_0^1 d\al\,e^{\al(1-\al)\tau}\,,\,\,\,\,\,\,\,\,\,\,
\,\,\,\,\,\,\,\,\,\,\, \tau=-s\na^2\,.
$$
The integral over the proper time $s$ is divergent and must be
regularized. Below we adopt the dimensional regularization in 
the form suggested by Brown and Cassidy \cite{brocas} (see also
\cite{bavi} for useful technical details). Also, inserting the 
$\exp(-sm^2)$ factor into (\ref{heat}) enables one to study the
UV limit $\tau/sm^2 \gg 1$ and the IR limit $\tau/sm^2 \ll 1$ 
instead of the limits $\tau \gg 1$ and $\tau \ll 1$ which
were investigated in \cite{bavi2}.  The interface between 
UV and IR limits is our main subject of interest here.

It proves useful to change the variable $sm^2=t$ 
and also denote 
$$
u=\frac{\tau}{t}=-\frac{\na^2}{m^2}\,.
$$
After simple calculations we arrive at the following 
representation for the effective action (\ref{effective}):
$$
{\bar \Ga}^{(1)}\,=\,-\,\frac{1}{2(4\pi)^2}\,\int d^4x g^{1/2}\,
\left(\frac{m^2}{4\pi \mu^2}\right)^{w-2}\int_{0}^\infty
dt\,e^{-t}\,\left\{\,
\frac{m^4}{t^{w+1}}\,+\,\Big(\xi-\frac16\Big)\,\frac{R\,m^2}{t^w}
\,+
\right.
$$
\beq
\left.
+\,\sum_{i=1}^{3}\,l^*_i\cdot R_{\mu\nu} M_i R^{\mu\nu}
\,+\,\sum_{j=1}^{5}\,l_j\cdot R M_j R\,\right\}\,,
\label{intermediate}
\eeq
where 
$$
l_1^*=1\,,\,\,\,\,\,\,\,\,\,\,\,\,\,\,\,\,
l_2^*=\frac16\,,\,\,\,\,\,\,\,\,\,\,\,\,\,\,\,\,
l_3^*=-1\,;
$$
\vskip 1mm
$$
l_1=\frac{1}{288}-\frac{1}{12}\,\Big(\xi-\frac16\Big)
+\frac{1}{2}\,\Big(\xi-\frac16\Big)^2
\,,\,\,\,\,\,\,\,\,\,\,\,
l_2=\frac{1}{24}-\frac{1}{2}\,\Big(\xi-\frac16\Big)\,,
$$
\vskip 1mm
$$
l_3=-\frac{1}{8}
\,,\,\,\,\,\,\,\,\,\,\,\,
l_4=-\frac{1}{16}+\frac{1}{2}\,\Big(\xi-\frac16\Big)
\,,\,\,\,\,\,\,\,\,\,\,\,
l_5=\frac{1}{8}
$$
and 
\beq
M_1=\frac{f(tu)}{u^2t^{w+1}}\,,\,\,\,\,\,\,\,\,\,
M_2=\frac{f(tu)}{ut^w}\,\,,\,\,\,\,\,\,\,\,\,
M_3=\frac{f(tu)}{u^2t^{w+1}}\,,\,\,\,\,\,\,\,\,\,
M_4=\frac{1}{ut^w}\,,\,\,\,\,\,\,\,\,\,
M_5=\frac{1}{u^2t^{w+1}}\,.
\label{M}
\eeq
The calculation of the integrals in (\ref{intermediate})
is quite tedious. Let us present just the result, using 
notations 
\beq
A =-\frac12\int_0^1 d\al\ln\Big[1+\al(1-\al)u\Big]
=1-\frac{1}{a}\ln \frac{1+a/2}{1-a/2}\,,
\,\,\,\,\,\,\,\,\,\,\,\,\,\
a^2=\frac{4\na^2}{\na^2-4m^2}
\label{A}
\eeq
(indeed, this is equivalent to the previous definition 
Eq. (\ref{first A})), relations (\ref{gamma}) and 
the expansion (we omit, systematically, those 
terms which contribute only to ${\cal O}(w-2)$-terms) 
$$
\left(\frac{m^2}{4\pi \mu^2}\right)^{w-2}
\,=\,1\,+\,(2-w)\ln \left(\frac{4\pi \mu^2}{m^2}\right) 
\,+\, ...\,.
$$
The effective action can be cast into the form 
$$
{\bar \Ga}^{(1)}
\,=\,\frac{1}{2(4\pi)^2}\,\int d^4x \,g^{1/2}\,
\left\{\,\frac{m^4}{2}\cdot\Big[\frac{1}{2-w}
+\ln \Big(\frac{4\pi \mu^2}{m^2}\Big)+\frac32\Big]\,+
\right.
$$$$
\left.
+\,\Big(\xi-\frac16\Big)\,m^2R\,
\Big[\,\frac{1}{2-w}
+\ln \Big(\frac{4\pi \mu^2}{m^2}\Big)+1\,\Big]\,+
\right.
$$$$
\left.
+\,\frac12\,C_{\mu\nu\al\be} \,\Big[\,\frac{1}{60\,(2-w)}
\,+\,\frac{1}{60}\ln \Big(\frac{4\pi \mu^2}{m^2}\Big)+k_W(a)
\,\Big] C^{\mu\nu\al\be}\,+
\right.
$$
\beq
\left.
+\,R \,\Big[\,\frac12\,\Big(\xi-\frac16\Big)^2\,
\left(\,\frac{1}{2-w}
+ \ln \Big(\frac{4\pi \mu^2}{m^2}\Big)\,\right)
+ k_R(a)\,\Big]\,R\,\right\}\,,
\label{final}
\eeq
where 
$$
k_W(a)\, = \,\frac{8A}{15\,a^4}
\,+\,\frac{2}{45\,a^2}\,+\,\frac{1}{150}\,,
$$
\beq
k_R(a)\, =
\,A\Big(\xi-\frac16\Big)^2-\frac{A}{6}\,\Big(\xi-\frac16\Big)
+\frac{2A}{3a^2}\,\Big(\xi-\frac16\Big)
+\frac{A}{9a^4}-\frac{A}{18a^2}+\frac{A}{144}+
\nonumber
\\
+\frac{1}{108\,a^2}
-\frac{7}{2160} + \frac{1}{18}\,\Big(\xi-\frac16\Big)^2\,.
\label{C}
\eeq
In Eq. (\ref{final}) we used the relation
$\,C^2=E+2W\,,$ that was justified in \cite{bavi2}, including 
the terms with non-local $1/\na^2$-type insertions.

It is easy to see that the above expression for the effective 
action perfectly corresponds to the polarization operator 
derived in the previous section. This coincidence represents 
reliable verification of the results and enables one to
obtain robust conclusions concerning the renormalization 
group in curved space-time. 

Another important check of the expression (\ref{final})
is the following. In the massless limit $m\to 0$ we expect 
to meet the standard massless result, and in particular 
the anomaly-induced effective action. Taking $m\to 0$ we 
arrive at the singular expression for $A$ in (\ref{A}). On the 
other hand, the consistency of the conformal theory (and in 
particular the existence of the anomaly-induced effective
action \cite{rei}) requires $\xi=1/6$. After we set 
$\xi=1/6$, the divergent $A$-dependent terms cancel, and in 
the $R[...]R$ sector we meet only the local term 
\beq
{\bar \Ga}^{(1)}(\xi=1/6,\,m\to 0)
\,=\,-\,\frac{1}{12\cdot 180(4\pi)^2}\,
\int d^4x \,g^{1/2}\,R^2 \,+...\,.
\label{anomaly-induced}
\eeq
Taking into account the well-known relation
$$
\frac{2}{\,g^{1/2}}\,g_{\mu\nu}
\frac{\de}{\de g_{\mu\nu}}\,\int d^4x \,g^{1/2}\,R^2 
\,=\,12\,\na^2R\,,
$$
we obtain the standard expression for the $\,\,\na^2R$-term
in the conformal anomaly 
\beq
 <T_\mu^\mu> \,\,=\,\,-\,\, \frac{1}{180\,(4\pi)^2}\,\na^2R
\,+\,...\,\,,
\label{anomaly}
\eeq
where the dots stand for the $\,E\,$ and $\,C^2\,$ terms 
which we do not discuss here. Let us remark that this test 
of the calculations is rather complete, because the coefficient 
in (\ref{anomaly}) depends on almost all terms in the expression 
for the effective action. Those terms which do not contribute 
to (\ref{anomaly}) directly, are verified through the cancelation 
of the singular (at $m\to 0$) structures. 

%%%%%%%%%%%%%%%%%%%%%%% / * \ %%%%%%%%%%%%%%%%%%%%%%%%%
 %%%%%%%%%%%%%%%%%%%%% /*%@%*\ %%%%%%%%%%%%%%%%%%%%%%%
 %%%%%%%%%%%%%%%%%%%%% \*%@%*/ %%%%%%%%%%%%%%%%%%%%%%%
%%%%%%%%%%%%%%%%%%%%%%% \ * / %%%%%%%%%%%%%%%%%%%%%%%

\section{Renormalization group equations}
\label{sec.4} 

In the $\overline{\rm MS}$ scheme the $\be$-function
of the effective charge $C$ is defined as 
\beq
\be_C(\overline{\rm MS})\,=\,\lim_{n\to 4}\,\mu\,\frac{dC}{d\mu}\,.
\label{beta}
\eeq
It is easy to see that when this procedure is applied to the 
expressions (\ref{final}) or (\ref{theta}), the 
$\,\,\be(\overline{\rm MS})$-functions for the parameters 
$\,\,{\La}/{G}$, $\,\,1/{G}$, $\,\,a_{1,2}\,\,$
are exactly the same as the usual ones, obtained in the 
covariant approach (\ref{group}) and (\ref{group1}). 

The disadvantage of the $\overline{\rm MS}$ scheme is that 
one can not control the decoupling of the massive quantum 
fields in an external gravitational field, hence one has to 
go beyond this scheme. The explicit calculations of the 
one-loop diagrams of massive fields presented in the 
previous sections enable one to apply the physical 
mass-dependent scheme of renormalization and raise a hope 
to observe the decoupling. Unfortunately, this approach is 
not universal, because the calculations have been performed 
through the perturbative expansion of the metric on the 
flat background. Also, the covariant calculation in section 
3 has been performed in the second order in curvature 
approximation and can not provide an information beyond that 
obtained in section 2. Therefore, one can not be sure to 
obtain a reliable information about the non-perturbative 
effects. At the same time, we can always use the correspondence 
with the standard results in the $\overline{\rm MS}$ scheme 
in the UV regime as a criterion of correctness of the results 
obtained in the mass-dependent scheme. 

The derivation of the $\be$-functions in the mass-dependent 
scheme has been described, e.g. in \cite{manohar}. Starting 
from the polarization operator, one has to subtract the 
counterterm at the momentum $p^2=M^2$, where $M$ is the 
renormalization point. Then, the $\be$-function is defined 
(instead of Eq. (\ref{beta})) as
\beq
\be_C = \lim_{n\to 4}\,M\,\frac{dC}{dM}\,.
\label{beta-mass-M}
\eeq
Mathematically, this is equivalent to taking the derivative 
$\,-pd/dp\,$ of the formfactors in the polarization operator. 
If we apply this procedure to the formfactor of 
the $C_{\mu\nu\al\be}^2$-term, the result for the $\be$-function
in a mass-dependent scheme is
\beq
\be_1\,=\, -\,\frac{1}{(4\pi)^2}\, \Big(\, 
\frac{1}{18a^2}\,-\,\frac{1}{180}\,-\, 
\frac{a^2-4}{6a^4}\,A\,\Big)\,,
\label{beta-mass}
\eeq
that is the general result for the one-loop $\be$-function
valid at any scale.
As the special cases we meet the UV limit $p^2 \gg m^2$
\beq
\be_1^{UV} \,=\,-\,\frac{1}{(4\pi)^2}\,\frac{1}{120} 
+ {\cal O}\Big(\frac{m^2}{p^2}\Big)\,,
\label{beta-UV}
\eeq
that agrees with the $\overline{\rm MS}$-scheme result (\ref{group1})
and also the IR limit $p^2 \ll m^2$: 
\beq
\be_1^{IR} \,=\,-\, \frac{1}{1680\,(4\pi)^2}\,\cdot\,
\frac{p^2}{m^2} \,\,
+ \,\,{\cal O}\Big(\frac{p^4}{m^4}\Big)\,.
\label{beta-IR}
\eeq

Similar calculations for the coefficient of the $R^2$-term
give 
$$
\be_2 \,\,=\,\, \lim_{n\to 4}\,M\,\frac{da_2}{dM} 
\,\,=\,\, -\,\frac{1}{(4\pi)^2}\, \Big\{\,
\frac{1}{8}\,(\,4A\, -\,a^2A\, + \,a^2\,)\,\Big(\xi-\frac16\Big)^2\,-
$$
\vskip 0.5mm
$$
-\,\,\frac{a^2-4}{48}\,\Big(\xi-\frac16\Big)
\,\,+\,\,\frac{(a^2-4)\,(a^2-12)A}{48\,a^2}\,\Big(\xi-\frac16\Big)\,\,+
$$
\beq
+\,\frac{a^2-4}{8}\,
\Big[\,\frac{A}{6a^2}-\frac{A}{144}-\frac{5\,A}{9a^4}
+\frac{1}{144}-\frac{5}{108a^2}\,\Big]\,\Big\}\,.
\label{beta-2 mass}
\eeq
In the UV limit $p^2 \gg m^2$ the $\be$-function is
(in agreement with the standard result (\ref{group1}))
\beq
\be_2^{UV}\,\,=\,\,-\,\frac{1}{2(4\pi)^2}\,\Big(\xi-\frac16\Big)^2 
+ {\cal O}\Big(\frac{m^2}{p^2}\Big)\,.
\label{beta 2-UV}
\eeq
while in the IR limit $p^2 \ll m^2$ we obtain: 
\beq
\be_2^{IR} \,\,=\,\,-\, \frac{1}{12\,(4\pi)^2} 
\,\,\Big[\,\Big(\xi-\frac16\Big)^2
-\frac{1}{15}\,\Big(\xi-\frac16\Big)
+\frac{1}{630}\,\Big]\,\cdot\,\frac{p^2}{m^2}
\,\,+\,\, {\cal O}\Big(\frac{p^4}{m^4}\Big)\,.
\label{beta 2-IR}
\eeq

The expressions (\ref{beta-mass})-(\ref{beta 2-IR}) are the 
main results of our work. In particular, the interpolation 
between the $\be$-functions in the high-energy regime 
(\ref{beta-UV}), (\ref{beta 2-UV}) and the $\be$-functions 
in the low-energy regime (\ref{beta-IR}), (\ref{beta 2-IR})
exhibits the decoupling phenomenon in the vacuum quantum effects
of the massive scalar particle. Let us remark that the 
correctness of the calculations of the $\be$-functions is
verified by: i) precise correspondence in the UV with the known 
expressions obtained in the $\overline{\rm MS}$ scheme; 
ii) great amount of cancelations of the singular ${\cal O}(m^4/p^4)$, 
${\cal O}(m^2/p^2)$ and constant ${\cal O}(1)$-terms 
in the expressions for $\be_{1,2}(IR)$ in (\ref{beta-IR})
and (\ref{beta 2-IR}); iii) correct result for the anomalous 
$\int R^2$-term in the massless limit. We remark that,
due to covariance of the effective action, the expressions 
for the $\be$-functions (\ref{beta-mass}) and 
(\ref{beta-2 mass}) should apply not only to the 2-point function
but also to the vertex corrections. 

At the same time, taking the derivative with respect to momenta 
of those terms in the effective action, which correspond to 
the cosmological constant and Einstein-Hilbert term, brings
zero result. Therefore, at this level we do not have an 
explicit interpretation for the renormalization group equations 
for $1/G$ and $\La/G$. Unfortunately, these $\be$-functions can 
not be calculated in the mass-dependent scheme through the 
perturbative expansion of the metric on flat background. 

%%%%%%%%%%%%%%%%%%%%%%%%%%%%%%%%%%%%%%%%%%%%%%%%%%%%%%%%%%%%%%%%%%%%%%%%
%%%%%%%%%%%%%%%%%%%%%%%%%%%%%%%%%%%%%%%%%%%%%%%%%%%%%%%%%%%%%%%%%%%%%%%%
%%%%%%%%%%%%%%%%%%%%%%%%%%%%%%%%%%%%%%%%%%%%%%%%%%%%%%%%%%%%%%%%%%%%%%%%
%%%%%%%%%%%%%%%%%%%%%%%%%%%%%%%%%%%%%%%%%%%%%%%%%%%%%%%%%%%%%%%%%%%%%%%%

%%% 6
\section{Discussions and conclusions}
\label{sec.5} 

Making the perturbative calculations in a physical 
mass-dependent
scheme, we arrived at the explicit expressions for the 
$\be$-functions for the coefficients of the higher-derivative 
terms $a_1$ and $a_2$. The decoupling of massive degrees 
of freedom proceeds according to the rule 
\beq
\be(IR)\,\,\, \sim \,\,\,const\,\cdot\, 
\Big(\frac{p^2}{m^2}\Big)\,,
\,\,\,\,\,\,\,\,\,\,\,\,\,\,\,\, p^2 < m^2\,,
\label{decoupling}
\eeq
which has been guessed in \cite{babic} for the cosmological 
constant by analogy with a similar phenomena in QED. However,
we have discovered this law not for the cosmological constant 
but only for the higher derivative terms.

As we have seen above, the perturbative calculations in the 
mass-dependent scheme do not reveal the $\be$-functions
in the case of  $1/G$ and $\La/G$. In these sectors
the UV limit of the mass-dependent scheme differs from 
that in the $\overline{\rm MS}$ scheme. The question is 
whether these $\be$-functions are really zero? From our 
point of view, this is not so. The reason for the difference
between two renormalization schemes is that actual calculations 
has been performed on the flat background, while the 
$\overline{\rm MS}$-based derivation is completely covariant. 

After all, there remains a possibility to have the 
scale dependence in the observables different from the 
polarization operator on the flat background. If this 
dependence really exists, it would be a non-perturbative 
effect with respect to the expansion (\ref{expand}), since 
the perturbative effects are controlled by  
general covariance. It is natural to expect that these 
non-perturbative effects really take place, for otherwise
there would not be correspondence between the 
mass-dependent and $\overline{\rm MS}$ schemes in the 
high-energy regime. However, it is not for sure that 
the decoupling law for these non-perturbative effects 
will be the same as for the perturbatively
accessible sectors such as the higher-derivative parameters
$\,\,a_{1,2}$.

It is interesting to analyze further the reasons why the 
"perturbative" $\be_{\La}$ and $\be_{1/G}$ are not 
seen in the mass-dependent scheme. In the perturbative
approach, as for any Quantum Field Theory in flat space-time, 
the $\be$-functions reflect the momentum-dependence 
of the formfactors in the polarization operator.
In the covariant formalism, this dependence corresponds
to the non-local terms in the effective action of vacuum.

%%%%%%%%%%%%%%%%%%%%%%%%%%%%%%%%%%%%%
%%%%%%%%%%%%%%%%%%%%%%%%%%%%%%%%%%%%%
%%%%%%%%%%%%%%%%%%%%%%%%%%%%%%%%%%%%%

We find it instructive once again to consider the QED example. 
Due to 
gauge invariance, momentum running of effective coupling constant 
in QED is connected with renormalization of the photon propagator 
(the $Z_3$ renormalization constant) and the effective action 

has the following non-local contribution (see, e.g.
\cite{Maroto}) in dimensional regularization for $p^2 \gg m_e^2$, 
\beq
S\, =\, -\frac{1}{4}F_{\mu\nu}F^{\mu\nu} 
+ \frac{e^2}{3(4\pi )^2}F_{\mu\nu}\,
\ln \Big(-\frac{\nabla^2}{\mu^2}\Big)\,F^{\mu\nu}.
\label{Z_3}
\eeq
This action can be regarded as a non-local version of the 
well-known Euler-Heisenberg Lagrangian \cite{EH}. 

Clearly, $\,\,\ln(-\nabla^2/\mu^2)\,\,$ corresponds to 
$\,\,\ln(p^2/\mu^2)\,\,$ contribution in momentum space, 
therefore, the electric charge `runs` as the momentum changes. 
Similarly in gravity, as follows from our Eq. (\ref{final}) 
in Sect. 3 for finite part there is
the following non-local contributions to the Euclidean action:
\beq
S\,=\,\frac{1}{2(4\pi)^2} \int d^4x \,g^{1/2} 
\,\Big[\,
\frac12\,C_{\mu\nu\al\be}\,k_W(a)\,C^{\mu\nu\al\be}
\,\, + \,\,R\,k_R(a)\,R\,\Big],
\label{finite}
\eeq
where the functions $k_W$ and $k_R$ are given by Eq. (\ref{C}). 
At high energy limit 
we find the familiar $\ln(-{\nabla^2}/{m^2})$-type 
contribution (see, e.g. \cite{buvo} where the high-energy
limit has been restored from the general covariant 
$\overline{\rm MS}$ renormalization group, and also
\cite{Maroto})
\beq
k_W \approx -\frac{1}{60}\,\ln \Big(-\frac{\nabla^2}{m^2}\Big)\,
\,\,\,\,\,\,\,\,\,\,\,\,\,{\rm and}\,\,\,\,\,\,\,\,\,\,\,\,\,\,
k_R \approx - \frac{1}{2} (\xi - \frac{1}{6})^2
\,\ln \Big(-\frac{\nabla^2}{m^2}\Big)\,.
\label{limit}
\eeq
At low energy massive particles decouple and $k_W$ and
$k_R$ can be formally 
expanded in series in $\nabla^2/m^2$ to give
$$
k_W \approx -\frac{1}{840}\,\Big(-\frac{\nabla^2}{m^2}\Big)\,,
$$
and
\beq
k_R \approx - \frac{1}{12}\,\Big[(\xi-\frac{1}{6})^2 
- \frac{1}{15}\,(\xi-\frac{1}{6}) + \frac{1}{630}\Big]
\,\Big(-\frac{\nabla^2}{m^2}\Big)\,.
\label{series}
\eeq
Thus, the IR running of the coupling constants $a_1$ and 
$a_2$ is quite similar to that of QED. However, for 
the cosmological constant and the Einstein term, the situation is 
different. The reason is simple. It is clear that unlike the
$C^2$ and $R^2$ terms (these two are similar to the 
$F_{\mu\nu}F^{\mu\nu}$ term in the case 
of QED) it is impossible to ensure renormalization group 
running for the cosmological 
constant and Einstein term by inserting a certain function of the 
operator ${\nabla^2}/{m^2}$ into the action terms because
this operation produces either zero (for the cosmological 
constant) or complete derivative (for the Einstein term).

According to \cite{nelspan, tmf, buchodin, book}, the 
renormalization group scaling 
in curved spacetime is defined as scaling of the metric 
$g_{\mu\nu} \to e^{2t}g_{\mu\nu}$. This is the key idea of 
renormalization group in 
curved spacetime because momentum is not defined in a general 
curved spacetime and, hence, the familiar flat space momentum 
scaling $p \to e^{-t}p$ cannot be used. The 
operator $\nabla^2=g^{\mu\nu}\nabla_{\mu}\nabla_{\nu}$ 
scales like $\nabla^2 \to e^{-2t}\nabla^2$ that corresponds to 
the flat spacetime scaling $p^2 \to e^{-2t}p^2$. 
If the running for the cosmological constant and Einstein 
terms could be seen in the framework of the perturbative
expansion, in the second order in $h_{\mu\nu}$ it can only 
be related with the terms 
$$
\int d^4x \,g^{1/2}
\,R_{\mu\nu}\,\Big(-\frac{m^2}{\nabla^2}\Big)^2\,
R^{\mu\nu}
\,,\,\,\,\,\,\,\,\,\,\,\,\,\,\,\,\,\,\,\,\,\,\,\,\,\,\,\,
\int d^4x\,g^{1/2}\,R\, \Big(-\frac{m^2}{\nabla^2}\Big)^2\,R
$$ 
for the cosmological constant and with the expressions
$$
\int d^4x\,g^{1/2}\,R_{\mu\nu}\, 
\Big(-\frac{m^2}{\nabla^2}\Big)\,
R^{\mu\nu}
\,,\,\,\,\,\,\,\,\,\,\,\,\,\,\,\,\,\,\,\,\,\,\,\,\,\,\,\,
\int d^4x\,g^{1/2}\,R\, \Big(-\frac{m^2}{\nabla^2}\Big)\,R
$$ 
for the Einstein-Hilbert term. Indeed, only these terms 
have an appropriate scaling, are of the second order 
in curvature, and admit the $\,\,\ln (-\nabla^2/m^2)\,\,$
insertion similar to that of (\ref{limit}). 

Within the perturbative scheme, one can 
see the renormalization group running of those terms
only, and not of the original cosmological and 
Einstein-Hilbert terms. But, our results show that the 
running for these two structures doesn't take place. 
Probably, this is related to the locality of the 
divergences in the effective action. Another obstacle for 
the appearance of these non-local terms arises in the IR 
limit, where the 
presence of $\nabla^2$ in denominator would violate a nice 
IR limit of gravity \cite{Weinberg,don}.
Indeed, the absence of the $\be$-functions for $1/G$ and 
$\La/G$ looks as an artefact of the perturbative expansion
in $h_{\mu\nu}$ rather than the fundamental property of the 
renormalization group in curved space-time. We expect that 
if the calculations were performed on a non-flat background,
there would be nontrivial $\be$-functions at high energy in 
a mass-dependent scheme. This would provide the 
correspondence with the covariant $\overline{\rm MS}$ result
(\ref{group}) at high energy and the information
concerning the decoupling at low energy. 

Another argument supporting the existence of the running
for $\La$ and $G$ comes
from the derivation of the effective action in the covariant
proper-time cut-off regularization (see Appendix B), 
where we have found the 
standard $\be$-functions for the cosmological and inverse 
Newton constants at the UV regime and the exponential 
decoupling of the massive fields contributions to these 
$\be$-functions in the IR regime (unfortunately, the 
IR limit in the covariant proper-time cut-off regularization
is not reliable).

In conclusion, we have performed the calculations 
in the mass-dependent renormalization scheme and 
found the explicit law for the decoupling of the 
massive degrees of freedom for the higher derivative
terms in the vacuum effective action. Only more
complicated calculations on the non-flat background can 
provide a reliable information about the renormalization
group for the cosmological and inverse Newton constants.

The form of the IR decoupling rule in the higher 
derivative sector indicates to the possibility of the 
soft transition between the stable and unstable regimes 
in the anomaly-induced inflation \cite{insusy,asta}. 
The detailed investigation of this application will be
reported separately. 

\acknowledgments 

One of the authors (I.Sh.) is grateful to M. Asorey and
J. Sola for discussions. 
E.G. acknowledges warm hospitality of the Physics Department
at the Federal University of Juiz de Fora. The work of the 
authors has been supported by the research 
grant from FAPEMIG (Minas Gerais, Brazil) and (I.Sh.) 
by the fellowship from CNPq (Brazil).

\vskip 12mm
\noindent
{\large\bf Note Added.} After this paper has been submitted,
we learned about the work by Bastianelli and Zirotti
\cite{basti}, which is also devoted to the derivation of 
quantum corrections for the massive non-minimal scalar. The 
calculations of \cite{basti} are based on the worldline 
formalism, and their final result corresponds to our 
intermediate expressions (\ref{fish}), (\ref{tadpole1})
before the integrals from the Appendix A are taken. 
Technically, this integration and the consequent 
identification with covariant terms in the gravitational 
action are the most difficult parts of our work. However, 
the correspondence of the intermediate formulas provides 
one more useful verification of the calculations.

%%%%%%%%%%%%%%%%%%%%%%% / * \ %%%%%%%%%%%%%%%%%%%%%%%%%
 %%%%%%%%%%%%%%%%%%%%% /*%|%*\ %%%%%%%%%%%%%%%%%%%%%%%
 %%%%%%%%%%%%%%%%%%%%% \*%|%*/ %%%%%%%%%%%%%%%%%%%%%%%
%%%%%%%%%%%%%%%%%%%%%%% \ * / %%%%%%%%%%%%%%%%%%%%%%%%%
$\,\,\,\,\,\,\,$

\vskip 8mm
%% \newpage
%%%%%%%%%%%%%%%%%%%%%%%%%%%%%%%%%%%%%%%%%%%%%%%%%%%%%%
%%%%%%%%%%%%%%%%%%%%%%%%%%%%%%%%%%%%%%%%%%%%%%%%%%%%%%
\section{Appendix A. The table of useful integrals}
\label{sec.6} 

In section 2 we have used dimensional regularization.
The following integrals were relevant:
$$
I_1\,=\,\int \frac{d^{2w}k}{(2\pi)^{2w}}
\,\frac{1}{(k^2+m^2)([k-p]^2+m^2)}
\,=\,\frac{{\Ga}(2-w)}{(4\pi)^w}\,(m^2)^{w-2}\,\,
_2F_1\Big(2-w,1,\frac32\,\,;\,\frac{-p^2}{4m^2}\Big)\,,
\eqno(A1)
$$
\vskip 1mm
$$
\int \frac{d^{2w}k}{(2\pi)^{2w}}
\,\frac{k^\mu}{(k^2+m^2)([k-p]^2+m^2)}
\,=\,\frac{I_1}{2}\,p^\mu\,=\,p^\mu I_2\,,
\eqno(A2)
$$
\vskip 1mm
$$
\int \frac{d^{2w}k}{(2\pi)^{2w}}
\,\frac{k^\mu k^\nu}{(k^2+m^2)([k-p]^2+m^2)}
\,=\,I_3\,\eta^{\mu\nu}\,+\,I_4\,p^\mu p^\nu\,,
\eqno(A3)
$$
\vskip 1mm
$$
I_3 \,=\,\frac{{\Ga}(1-w)}{2\,(4\pi)^w}\,(m^2)^{w-1}\,\,
_2F_1\Big(1-w,1,\frac32\,\,;\,\frac{-p^2}{4m^2}\Big)\,,
\eqno(A4)
$$
\vskip 1mm
$$
I_4 \,=\,\frac{{\Ga}(2-w)}{2\,(4\pi)^w}\,(m^2)^{w-2}\,
\left[\,
_2F_1\Big(2-w,1,\frac32\,\,;\,\frac{-p^2}{4m^2}\Big)
\,-\,\frac13\,_2F_1\Big(2-w,2,\frac52\,\,;\,\frac{-p^2}{4m^2}\Big)
\,\right]\,,
\eqno(A5)
$$
\vskip 1mm
$$
\int \frac{d^{2w}k}{(2\pi)^{2w}}
\,\frac{k^\mu k^\nu k^\al}{(k^2+m^2)([k-p]^2+m^2)}
\,=\,I_5\,\,p^\mu p^\nu p^\al \,+\,
3\,p^{(\al} \eta^{\mu\nu)}\,I_6\,,
\eqno(A6)
$$
where the parenthesis indicate to the symmetrization
(e.g. $D^{(\al\be)}=1/2[D^{\al\be}+D^{\be\al}]$),
\vskip 1mm
$$
I_5 \,=\,\frac{{\Ga}(2-w)}{2\,(4\pi)^w}\,(m^2)^{w-1}\,\,
\left[\,
_2F_1\Big(2-w,1,\frac32\,\,;\,\frac{-p^2}{4m^2}\Big)
\,-\,
\frac12\,\,_2F_1\Big(2-w,2,\frac52\,\,;\,\frac{-p^2}{4m^2}\Big)
\,\right]\,,
\eqno(A7)
$$
\vskip 1mm
$$
I_6 \,=\,\frac{{\Ga}(1-w)}{4\,(4\pi)^w}\,(m^2)^{w-2}\,
_2F_1\Big(1-w,1,\frac32\,\,;\,\frac{-p^2}{4m^2}\Big)\,,
\eqno(A8)
$$
\vskip 1mm
$$
\int \frac{d^{2w}k}{(2\pi)^{2w}}
\,\frac{k^\mu k^\nu k^\al k^\be}{(k^2+m^2)([k-p]^2+m^2)}
\,=\,I_7\,\,p^\mu p^\nu p^\al p^\be
\,+\,6\,p^{(\al} \eta^{\mu\nu}p^{\be)}I_8
\,+\,3\,\eta^{(\al\be} \eta^{\mu\nu)}\,I_9\,,
\eqno(A9)
$$
\vskip 1mm
$$
I_7 \,=\,\frac{{\Ga}(2-w)}{(4\pi)^w}\,(m^2)^{w-2}\,\,
\left[\,
\frac12\,\,_2F_1\Big(2-w,1,\frac32\,\,;\,\frac{-p^2}{4m^2}\Big)
\,-\,
\frac13\,\,_2F_1\Big(2-w,2,\frac52\,\,;\,\frac{-p^2}{4m^2}\Big)\,+
\right.
$$$$
\left.
\,+\,
\frac{1}{30}\,\,_2F_1\Big(2-w,3,\frac72\,\,;\,\frac{-p^2}{4m^2}\Big)
\,\right]\,,
\eqno(A10)
$$
\vskip 1mm
$$
I_8 \,=\,\frac{{\Ga}(1-w)}{4(4\pi)^w}\,(m^2)^{w-1}\,\,
\left[\,_2F_1\Big(1-w,1,\frac32\,\,;\,\frac{-p^2}{4m^2}\Big)
\,-\,
\frac13\,\,_2F_1\Big(1-w,2,\frac52\,\,;\,\frac{-p^2}{4m^2}\Big)
\,\right]\,,
\eqno(A11)
$$
$$
I_9 \,=\,\frac{{\Ga}(-w)}{4(4\pi)^w}\,(m^2)^{w}\,\,
_2F_1\Big(-w,1,\frac32\,\,;\,\frac{-p^2}{4m^2}\Big)\,.
\eqno(A12)
$$

Here, 
$$
_2F_1\Big(a,b,c \,;z\Big)\,=\,\sum_{n=0}^{\infty}\,
\frac{(a)_n\,(b)_n}{(c)_n}\,\frac{z^n}{n!}
\eqno(A13)
$$
is the hypergeometric function of $z$ and 
$$
(a)_n\,=\,a(a+1)...(a+n-1)\,=\,\frac{{\Ga}(a+n)}{{\Ga}(a)}
$$
is the Pochhammer symbol.

%% \newpage
%%%%%%%%%%%%%%%%%%%%%%%%%%%%%%%%%%%%%%%%%%%%%%%%%%%%%%
%%%%%%%%%%%%%%%%%%%%%%%%%%%%%%%%%%%%%%%%%%%%%%%%%%%%%%
\section{Appendix B. 
The covariant proper-time cut-off regularization}
\label{sec.7} 

In order to understand better our result concerning the 
$\,\be$-functions, let us perform an alternative covariant 
calculation using the cut-off regularization of the 
proper-time integral in the Schwinger-De Witt approach.
We shall use the same proper-time integral and the expression 
for the heat-kernel as in the section 3. In some sense, the 
use of the covariant cut-off regularization provides a small 
advantage, because one can investigate 
not only logarithmic, but also quadratic and quartic 
divergences. 

In the covariant cut-off regularization the effective
action is defined as
$$
{\bar \Ga}^{(1)}\,
=\,\,-\,\frac12\,\int\limits_{1/\Om^2}^{\infty}\,
\frac{ds}{s}\,\tr\,K(s)\,,
\eqno(B1)
$$
where $\Om$ is a cut-off parameter with the dimension of
mass and all other notations are the same as in Eq. 
(\ref{heat}). Let us perform the analysis separately 
for the cosmological 
constant, Einstein-Hilbert terms and for the 
higher-derivative terms.   
\vskip 2mm

i) For the cosmological constant term, using the same 
change of variables as in section 3, we obtain 
$$
\Ga_{\La}^{(1)}\,=
\,-\,\frac12\int\limits_{1/\Om^2}^{\infty}\,\frac{ds}{s}\,
e^{-m^2s}\,\frac{1}{(4\pi s)^2}
\,=\,-\,\frac{1}{2(4\pi)^2}
\int\limits_{m^2/\Om^2}^{\infty}\,dt\,t^{-3}\,
e^{-t}\,=\,-\,\frac{m^4}{2(4\pi)^2}\,
\Ga\Big(-2,\frac{m^2}{\Om^2}\Big)\,,
\eqno(B2)
$$
where
$$
\Ga(\al,x)\,=\,\int\limits_{x}^{\infty}\,dt\,e^{-t}\,t^{\al - 1}
\eqno(B3)
$$
is the incomplete gamma function (see, e.g. \cite{GR}, formula
8.350.2). Using the expansion
$$
\Ga(-n,x)\,=\,\frac{(-1)^n}{n!}\,\Big[\,\Ga(0,x)\,\,
-\,\,e^{-x}\,\sum_{m=0}^{n-1}\,\frac{(-1)^m \,m!}{x^{m+1}}\,
\Big]\,,
\eqno(B4)
$$
where $\,\,\,\,\Ga(0,x)=-{\rm Ei}(-x)\,\,\,\,$ and
$$
{\rm Ei}(-x)\,=\,C\,+\,\ln(x)
\,+\,\sum\limits_{k=1}^{\infty}\,\frac{(-x)^k}{k\cdot k!}\,,
\,\,\,\,\,\,\,\,\,\,(x>0)\,,
\eqno(B5)
$$
for the large cut-off limit $\,\Om^2\gg m^2\,$ we find
$$
\Ga_{\La}^{(1)}\,\approx\,-\,\frac{1}{4\,(4\pi)^2}\,
\Big[\,\Om^4 \,+\, 2\Om^2m^2 
\,+\, m^4\,\ln \Big(\frac{\Om^2}{m^2}\Big)\,\Big]\,+\,...\,.
\eqno(B6)
$$
The structure of divergences is the standard one. The 
divergences can be canceled by the corresponding 
counterterms. Let us choose them in the form 
$$
\De S \,=\,+\,\frac{1}{4\,(4\pi)^2}\,
\Big[\,(\Om^4 + k_1\mu^4)\,+\, 2(\Om^2+ k_2\mu^2)m^2 
\,+\, m^4\,\ln \Big(\frac{\Om^2}{\mu^2}\Big)\,\Big]\,.
\eqno(B7)
$$
If the normalization condition is chosen in a simplest 
and natural way $\,k_1=k_2=0$,
the finite part of the effective action will not depend 
on the quartic and quadratic divergences (see also 
discussion in \cite{nova}). It is easy to see that the 
$\,\,\be_\La\,\,$ can be calculated using the formula
(\ref{group}) and we arrive at the usual 
$\overline{\rm MS}$-scheme renormalization group equation 
for the cosmological constant. Indeed, the same 
$\be$-function can be obtained by taking the logarithmic 
derivative of the corresponding term in (B6) with respect 
to the covariant 
cut-off $\,\,\Om\,d/d\Om$ with negative sign. Below we 
shall neglect the non-logarithmic divergences and 
always define $\,\be$-functions by taking the last type of  
derivative. 

In order to see the decoupling of the massive field at 
low energy, one can try to consider another extreme case 
$\Om^2 \ll m^2$. The solution (B2) gives, after 
we apply to it the derivative $\,\Om (d / d\Om)$
$$
\be_\La^{IR}\,=\,-\,\frac{\Om^4}{(4\pi)^2}\,e^{-m^2/\Om^2}
\eqno(B8)
$$
Another way of getting the same formula is to use the 
asymptotics of the incomplete gamma function
(see \cite{GR}, formula 8.357) for large $|x|$
$$
\Ga(\al,x)\,=\,x^{\al-1}\,e^{-x}\,\Big[\,
\sum_{m=0}^{M-1}\frac{(-1)^m\,\Ga(1-\al+m)}{x^m\,\Ga(1-\al)}
\,\,+\,\,{\cal O} (|x|^{-M})\,\Big]\,,\,\,\,\,\,\,\,\,\,\,\,
|x|\to \infty
\eqno(B9)
$$
and then apply the derivative $\,\Om (d / d\Om)$
to the expression
$$
-\,\frac{m^4}{2(4\pi)^2}\,\Ga\Big(-2,\frac{m^2}{\Om^2}\Big)
\,\approx\,
-\,\frac{m^4}{2(4\pi)^2}\,\frac{\Om^6}{m^6}\,e^{-m^2/\Om^2}\,.
\eqno(B10)
$$

The expression (B8) can be understood as a 
very fast decrease of $\be_\La$ in the region below the natural 
scale $\,m^2$. It is tentative to consider Eq. 
(B10) as a hint supporting the sharp cut-off decoupling 
for the cosmological constant \cite{nova}, in contrast to 
the hypothesis $\,\be_\La(IR) \sim \mu^2/m^2\cdot \be_\La(UV)\,$ 
which has 
been suggested in \cite{babic}. Unfortunately, the physical 
sense of the cut-off in the gravitational setting 
and especially the mathematical consistency of the 
$\Om^2 \ll m^2$ limit for the heat-kernel representation  
are not obvious. Moreover, for the higher derivative sector 
this definition is, as it will be shown below, 
contradictory. Finally, the decoupling in the lower-derivative 
sector of the gravitational action can not be understood by 
using the covariant proper-time cut-off, and we present it 
here just to complete the discussion. 

%%%%%%%%%%%%%%%%%%%%%%%%%%%%%%%%%%%%%%%%%%%%%%%%%%%%%%%%
\vskip 2mm

ii) The calculation of the divergences and the 
renormalization of the Einstein-Hilbert term goes the 
same way as for the cosmological constant term. 
The effective action is 

$$
\Ga_{EH}^{(1)}
\,=\,\frac12\,\Big(\xi-\frac16\Big)\,
\int\limits_{1/\Om^2}^{\infty}\,\frac{ds}{(4\pi s)^2}\,
e^{-m^2s}\,R\,=
$$
$$
=\,\,\,\frac{m^2R}{2(4\pi)^2}\,
\Big(\xi-\frac16\Big)\int\limits_{m^2/\Om^2}^{\infty}
\,\frac{dt}{t^2}\,e^{-t}
\,\,\,=\,\,\,\frac{m^2\,R}{2(4\pi)^2}\,\Big(\xi-\frac16\Big)
\,\Ga\Big(-1,\frac{m^2}{\Om^2}\Big)\,.
\eqno(B11)
$$
For $\,\Om^2\gg m^2\,$ the result is
$$
\Ga_{EH}^{(1)}\, \approx \,\frac{1}{2\,(4\pi)^2}\,
\Big(\xi-\frac16\Big)\,R\,\Big[\,\Om^2 
\,-\, m^2\,\ln \Big(\frac{\Om^2}{m^2}\Big)\,\Big]
\,+\,...\,.
\eqno(B12)
$$
The form of the counterterms and the expression for the 
$\be$-function is obvious. 
$$
\be_R^{UV}
\,\,\,=\,\,\,\frac{m^2}{2(4\pi)^2}\,\Big(\xi-\frac16\Big)\,.
\label{beta R in UV}
\eqno(B13)
$$
This result agrees with the previous one Eq. (\ref{group1}) 
obtained in the $\overline{\rm MS}$ scheme. In the low-energy limit 
$\,\Om^2\ll m^2\,$ we obtain
$$
\be_R^{IR}\,=\,-\,\frac{\Om^2}{(4\pi)^2}\,e^{-m^2/\Om^2}\,.
\eqno(B14)
$$
Of course, this is the same type of decoupling which we 
already met for the cosmological constant case. Unfortunately,
this IR limit is inconsistent, as we shall see in brief.

%%%%%%%%%%%%%%%%%%%%%%%%%%%%%%%%%%%%%%%%%%%%%%%%%%%%%%%%%%%%%
\vskip 2mm

iii) For the higher-derivative sector we consider, for the 
sake of simplicity, only the $R_{\mu\nu}(..)R^{\mu\nu}$-type
term. This term defines the renormalization of the $C^2$-terms,
and therefore we can expect the correspondence with the 
expression (\ref{beta-mass}) in both UV and IR limits. After 
the standard change 
of variables we arrive at the integral representation
$$
\Ga_{R_{\mu\nu}^2}^{(1)}\,=\,-\,\frac{1}{2(4\pi)^2}\,
\int\limits_{m^2/\Om^2}^{\infty}\,\frac{dt}{t}\,e^{-t}\,
R_{\mu\nu}\,\Big[\,\frac{1}{t^2u^2}\,\int_0^1 d\al\,
e^{-\al(1-\al)tu}\,-\,\frac{1}{t^2u^2}\,+\,\frac{1}{6tu}
\,\Big]\,R^{\mu\nu}\,,
\eqno(B15)
$$
where, as before, $u=-\na^2/m^2$.
After some algebra we arrive at the $\be$-function
$$
\be_1\,=\,-\,\frac{1}{2(4\pi)^2}\,
e^{-m^2/\Om^2}\,\Big[\,
\frac{\Om^4}{\na^4}\,\int_0^1 d\al\,
e^{\al(1-\al){\na^2}/{\Om^2}}\,-\,\frac{\Om^4}{\na^4}
\,-\,\frac{1}{6}\,\frac{\Om^2}{\na^2}\,\Big]\,.
\eqno(B16)
$$
At low energies we expand $\,\exp[{\al(1-\al){\na^2}/{\Om^2}}]\,$ 
into the series in ${\na^2}/{\Om^2}$ and arrive at the expression
$$
\be_1\, \approx \,-\,\frac{1}{120\,(4\pi)^2}\,e^{-m^2/\Om^2}\,.
\eqno(B17)
$$
This expressions is in disagreement with the previous one 
Eq. (\ref{beta-IR}), which was obtained in the framework 
of the standard methods and the physical mass-dependent 
scheme. Hence, the IR limit $\Om^2 \ll m^2$ 
of the renormalization group equations in the covariant cut-off
approach is contradictory. In fact, this contradiction shows,
once again, that the physical interpretation of the 
$\mu$-dependence in gravity is quite complicated. It is 
possible that 
this interpretation depends on the concrete choice of the 
background metric and can not be formulated in the general 
form.

The source of the difference between the expressions 
(\ref{beta-IR}) and (B17) is obvious. The proper-time 
integral representation of the effective action is 
formally consistent if the integration performs from 
zero to infinity (see, e.g. \cite{DeWitt75}).
Indeed, the lower limit of this integral is divergent 
and its UV regularization is related to the renormalization 
group. In particular, this can be seen from
the UV limit $\Om^2 \gg m^2$  which is consistent in all sectors. 
But, when taking the  $\Om^2 \ll m^2$ limit, we lose the 
link with the effective action and hence the corresponding
results do not have much sense.

%%%%%%%%%%%%%%%%%%%%%%%%%%%%%%%%%%%%%%%%%%%%%%%%%%%%%%%%%%%%%%%%%%%
%%%%%%%%%%%%%%%%%%%%%%%%%%%%%%%%%%%%%%%%%%%%%%%%%%%%%%%%%%%%%%%%%%%
%%%%%%%%%%%%%%%%%%%%%%%%%%%%%%%%%%%%%%%%%%%%%%%%%%%%%%%%%%%%%%%%%%%
%%%%%%%%%%%%%%%%%%%%%%%%%%%%%%%%%%%%%%%%%%%%%%%%%%%%%%%%%%%%%%%%%%%

%%%%%%%%%%%%%%%%%%%%%%%%%%%%%%%%%%%%%%%%%%%%%%%%%%%%%%%%%%%%%%%

\end{document}